\pdfoutput=1

\documentclass[11pt,twoside,a4paper,cmspaper,final,collab]{cms-tdr}
\def\svnVersion{375164}\def\cmsCernNoTag{CERN-PH-EP/2015-190}\def\cmsCernDate{\today}\def\cmsMessage{Published in Physics Letters B as \href{http://dx.doi.org/10.1016/j.physletb.2016.02.002}{\doi{10.1016/j.physletb.2016.02.002.}}}

\begin{document}\cmsNoteHeader{EXO-12-011}

\hyphenation{had-ron-i-za-tion}
\hyphenation{cal-or-i-me-ter}
\hyphenation{de-vices}
\RCS$Revision: 315817 $
\RCS$HeadURL: svn+ssh://svn.cern.ch/reps/tdr2/papers/EXO-12-011/trunk/EXO-12-011.tex $
\RCS$Id: EXO-12-011.tex 315817 2015-12-20 23:14:01Z alverson $
\pdfminorversion=5
\newlength\cmsFigWidth
\ifthenelse{\boolean{cms@external}}{\setlength\cmsFigWidth{0.85\columnwidth}}{\setlength\cmsFigWidth{0.4\textwidth}}
\ifthenelse{\boolean{cms@external}}{\providecommand{\cmsLeft}{top}\xspace}{\providecommand{\cmsLeft}{left}\xspace}
\ifthenelse{\boolean{cms@external}}{\providecommand{\cmsRight}{bottom}\xspace}{\providecommand{\cmsRight}{right}\xspace}

\providecommand{\PZ}{\ensuremath{\cmsSymbolFace{Z}}\xspace}
\renewcommand{\PW}{\ensuremath{\cmsSymbolFace{W}}\xspace}
\providecommand{\pTtau}{\ensuremath{\pt^{\PGt}}\xspace}
\providecommand\PWprTauNu{\ensuremath{\PWpr \to \PGt\nu}\xspace}
\providecommand{\MT}{\ensuremath{M_{\mathrm{T}}}\xspace}
\providecommand{\MTlower}{\ensuremath{\MT^\text{min}}\xspace}
\providecommand{\tauh}{\ensuremath{\tau_{\mathrm{h}}}\xspace}
\providecommand{\CL}{CL\xspace}
\newcolumntype{.}{D{.}{.}{-1}}

\cmsNoteHeader{EXO-12-011}
\title{\texorpdfstring{Search for \PWpr decaying to tau lepton and neutrino
in proton-proton collisions at $\sqrt{s}=8$\TeV}{Search for W' decaying to tau lepton and neutrino in proton-proton collisions at sqrt(s) = 8 TeV}}

\date{\today}

\abstract{
The first search for a heavy charged vector boson in the final state with a tau lepton and a neutrino is reported, using 19.7\fbinv
of LHC data at $\sqrt{s} = 8$\TeV.
A signal would appear as an excess of events with high transverse mass, where the standard model background is low.
No excess is observed.
Limits are set on a model in which the $\PWpr$ decays
preferentially to fermions of the third generation. These results substantially extend previous constraints on this model.
Masses below 2.0 to 2.7\TeV are excluded,
depending on the model parameters. In addition, the existence of a $\PWpr$ boson with universal fermion couplings is excluded at 95\%
confidence level, for $\PWpr$ masses below 2.7\TeV.
For further reinterpretation a model-independent limit on potential signals for various transverse mass thresholds is also presented.

}

\hypersetup{%
pdfauthor={CMS Collaboration},%
pdftitle={Search for W' decaying to tau lepton and neutrino in proton-proton collisions at sqrt(s) = 8 TeV},%
pdfsubject={CMS},%
pdfkeywords={CMS, physics, W' decays}}

\maketitle

\section{Introduction}
New heavy gauge bosons are predicted by various extensions of the standard model (SM).
Charged heavy gauge bosons are generally referred to as \PWpr
~\cite{reference-model}.
Non-universal gauge interaction models (NUGIM)~\cite{Chiang:2009kb,Li:1981nk,Muller:1996dj,Malkawi:1996fs}
predict a larger \PWpr-boson branching fraction to the third generation of fermions.
Searches for a \PWpr boson decaying to a tau lepton and neutrino have never been performed before, while
the electron and muon channels have been studied extensively
at the Tevatron~\cite{cdf-limit, d0-limit} and by the ATLAS and CMS experiments
at the LHC~\cite{ATLAS-paper-14,EXO-12-060}.
This Letter describes a search for a \PWpr boson decaying to a tau lepton and a neutrino
with the CMS detector~\cite{cms} at the CERN LHC, using proton-proton collisions collected in 2012 at a center-of-mass energy of 8\TeV.
The data set corresponds to an integrated luminosity of $19.7\pm0.5$\fbinv.
The results are interpreted in the context of the sequential standard model (SSM) \PWpr boson~\cite{reference-model} as well as an extended gauge group
NUGIM~\cite{Kim:2011qk,AlexLisaPaper,Chiang:2009kb}.
The signature of a \PWpr-boson event is similar to that of a \PW-boson event in which the \PW boson is produced ``off shell" with a high mass.
Events of interest are those in which the only detectable products of the \PWpr decay form a single hadronically
decaying tau (\tauh).
The hadronic decays of the tau lepton are experimentally distinctive because they result in low charged hadron multiplicity, unlike QCD jets, which have high hadron multiplicity, or other leptonic \PWpr decays, which have none.
In contrast, the decays $\PWpr \to \tau\nu_{\tau} \to \Pe \nu_{\Pe} \nu_{\tau}\nu_{\tau}$ and $\PWpr \to \tau
\nu_{\tau} \to \mu \nu_{\mu} \nu_{\tau} \nu_{\tau}$ cannot be distinguished from $\PWpr \to \Pe \nu_{\Pe}$ and
$\PWpr \to \mu \nu_{\mu}$, thus they suffer from low significance and are not selected in this analysis but rather
in the corresponding leptonic (\Pe,$\mu$) \PWpr searches.

\section{Physics Models}
In the SSM, the \PWpr boson is a heavy analogue of the \PW boson. It is a narrow resonance with fermionic decay modes and branching fractions similar to those of the SM \PW boson, with the addition of
the decay $\PWpr \to \PQt\PQb$, which becomes relevant for \PWpr-boson masses larger than 180\GeV.
If the \PWpr boson is heavy enough to decay to top and bottom quarks, the SSM branching fraction for the decay $\PWpr \to \tau\nu$ is 8.5\%. Under these assumptions, the total width of a 1\TeV \PWpr boson is about 33\GeV.
Decays of the \PWpr boson into $\PW\Z$ bosons depend on the specific model assumptions and are usually considered to be suppressed in the SSM, as assumed by the current search and by previous searches in other final states~\cite{EXO-12-060,Chatrchyan:2014koa}.
If the \PWpr interacts with left-handed particles and right-handed anti-particles ($V-A$ coupling),
interference with the SM W boson is expected~\cite{dudkov, criticalPaper, WprimeHelicity}.

\begin{figure}[htp]
\centering
\includegraphics[width=0.48\textwidth]{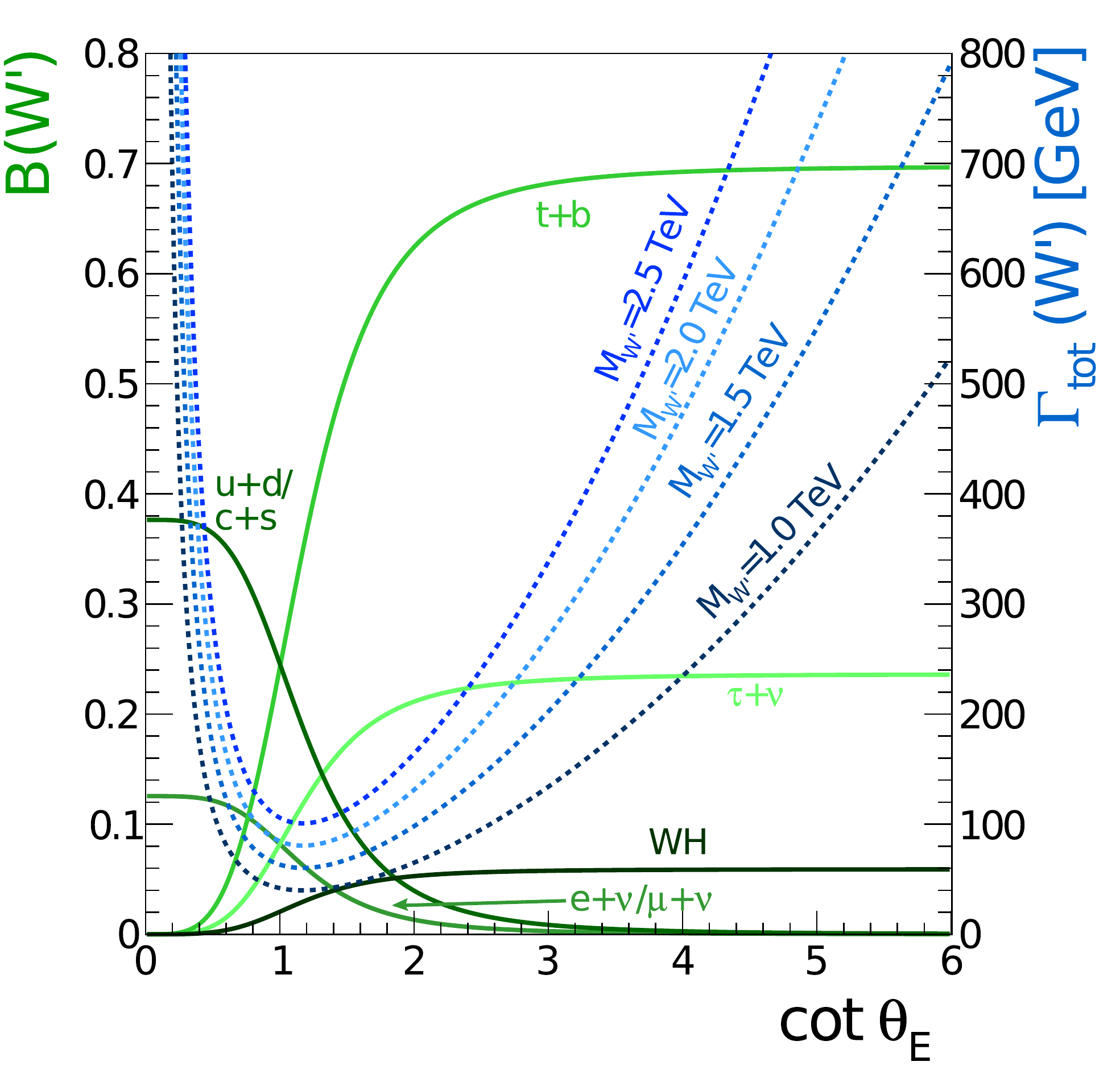}
\caption{\label{fig:BRPlot}
Branching fractions (left-hand scale and solid lines) and total width (right-hand scale and dotted lines) for \PWpr decays in the NUGIM, as calculated in
Refs.~\cite{Kim:2011qk,AlexLisaPaper,Chiang:2009kb}.
For $\cot{\theta_E} = 1$ the values are the same as those in the SSM, rescaled to accommodate the $\PW \PH$ decay channel.
}
\end{figure}

Models with non-universal couplings predict an enhanced branching fraction to the third generation of fermions and explain the large mass of the top quark.
In the other model studied in this analysis, NUGIM~\cite{Kim:2011qk,AlexLisaPaper,Chiang:2009kb},
the weak SM $SU(2)_{W}$ group is a low-energy limit of two gauge groups, a light $SU(2)_l$ and a heavy $SU(2)_h$, which couple only to the light fermions of the first two generations and to the heavy fermions of the third generation,
respectively. These two groups mix such that an SM-like $SU(2)_{\PW}$ and an extended group $SU(2)_E$ exist.
The second $SU(2)_E$ extended gauge group gives rise to additional gauge bosons such as a \PWpr.
The mixing of the two gauge groups is described by a mixing angle of the extended group $\theta_E$, which modifies the coupling to the heavy bosons.
Hence
the mixing changes the production cross section and, as illustrated in Fig.~\ref{fig:BRPlot},
the branching fractions of the \PWpr.
For $\cot \theta_E \gtrsim 3$ the \PWpr boson decays to fermions of the third generation only, whereas at $\cot \theta_E = 1$ the branching fractions are identical to those of the SSM, and the \PWpr
couples democratically to all fermions.
For $\cot \theta_E < 1$ the decays into light fermions are dominant. In the NUGIM, the decay into $\PW\Z$ bosons is negligible by construction.
In either the SSM or the NUGIM, the presence of a \PWpr-boson signal over the \PW-boson background could be observed in the distribution of the transverse mass (\MT) of the \tauh
and the missing transverse energy (\MET):
\begin{equation}
\MT = \sqrt{2 \, \pTtau  \, \MET \, [1 - \cos \Delta \phi(\tau,\ptvecmiss)] },
\label{eqn:mt}
\end{equation}
where \pTtau denotes the \pt of the \tauh and $\MET=\abs{\ptvecmiss}$, where \ptvecmiss is defined as ${-}\sum \ptvec$ of all reconstructed particles. The angle in the transverse plane between \ptvecmiss
and the direction of \tauh is denoted $\Delta
\phi(\tau,\ptvecmiss)$.

\section{Generation of Background and Signal Samples}
\label{sec:simulation}
The major SM backgrounds are dominated by \PW and \PZ+jets production and are generated using \MADGRAPH 5.1~\cite{madgraph} (for on-shell \PW and \PZ+jets backgrounds), \PYTHIA
6.426~\cite{Sjostrand:2006za}
(for off-shell \PW, $\PW\PW$, $\PW\PZ$, and $\PZ\PZ$ backgrounds) and \POWHEG 1.0~\cite{Nason:2004rx,Frixione:2007vw, Re:2010bp,Frederix:2012dh,Alioli:2011as} (for \ttbar and single \PQt{}+jets).
The tau decay is simulated by \TAUOLA~\cite{tauola} for all samples. For the hadronization of the \MADGRAPH background, \PYTHIA is used.
The response of these events in the CMS detector is simulated using \GEANTfour~\cite{geant4}.
The backgrounds are produced at leading-order (LO), but reweighted to higher order cross sections.
For the main \PW+jets background, a differential cross section as a function of the mass of the \PW-boson decay products is reweighted, incorporating
next-to-next-to-leading-order (NNLO) QCD and
next-to-leading-order (NLO) electroweak corrections. The effect with respect to the LO calculation corresponds to a K-factor of 1.3 at a mass of 0.3\TeV and drops for higher masses to 1.1 for a mass of 1\TeV.
The calculation uses Monte Carlo generators \textsc{mcsanc}~1.01~\cite{mcsanc} and \FEWZ~3.1~\cite{fewz}, following the recommended combination from
Ref.~\cite{leshouches2013SM}.
For the \PZ+jets background, the inclusive NNLO QCD cross section is calculated using $\FEWZ$.
For \ttbar events, the inclusive NNLO calculation from~\cite{Czakon:2013goa} is used.
For the diboson ($VV$) backgrounds, inclusive NLO QCD cross sections are calculated using \MCFM~6.6~\cite{Campbell:2011bn}.
The background contribution from multijet events is estimated from control samples in data.
The signal events for the SSM \PWpr are generated with \PYTHIA with NNLO cross sections from \FEWZ.
The NUGIM signals are generated with \MADGRAPH 4.5.1~\cite{madgraph} and hadronized with \PYTHIA.
The parton distribution functions (PDFs) used are CTEQ6L1~\cite{Pumplin:2002vw} for leading order simulation and CTEQ10~\cite{Gao:2013xoa} for (N)NLO simulation. The electroweak NLO calculation NNPDF 2.3 at NNLO QCD with and without QED contributions~\cite{Ball:2013hta} are used.

\section{The CMS Detector}
The central feature of the CMS apparatus is a superconducting solenoid of 6\unit{m} internal diameter, providing a magnetic field of 3.8\unit{T}.
Within the solenoid volume are a silicon pixel and strip tracker, a lead tungstate crystal electromagnetic calorimeter (ECAL), and a brass and scintillator hadron calorimeter (HCAL), each composed of a barrel and two endcap sections. Muons are measured in gas-ionization detectors embedded in the steel flux-return yoke outside the solenoid. Extensive forward calorimetry complements the coverage provided by the barrel and endcap detectors.
A particle-flow (PF) event algorithm~\cite{CMS-PAS-PFT-10-001} is used to reconstruct the events, identify the tau candidates and determine the missing \ET.
The algorithm reconstructs and identifies single particles with an optimized combination of all subdetector information.
The events are triggered by the CMS trigger system, which is split into two levels, a first level (L1) composed of custom hardware processors, and a high-level trigger (HLT) processor farm.
For this analysis a ``jet plus \MET{}" trigger is used, with thresholds of $\pt > 80\GeV$ for the jet and $\MET > 105\GeV$, where the latter is seeded at L1 in the calorimeter with \MET above 40\GeV.
Both objects are reconstructed at the HLT level using the PF event reconstruction.
A more detailed description of the CMS detector,
together with a definition of the coordinate system used and the relevant kinematic variables,
can be found in Ref.~\cite{cms}.

\section{Reconstruction and Identification of Physics Objects}
\label{sec:reco}

Tau reconstruction in CMS~\cite{TAU-14-001} is applied to jets clustered from PF objects, using the anti-\kt algorithm with a parameter $R = 0.5$.
Tau candidates must be distinguished from quark or gluon jets (QCD jets in the following).
The hadronic tau decays, \tauh,
are reconstructed using the ``hadron-plus-strips" (HPS) algorithm, which is based on decay modes
proceeding via specific intermediate resonances, with a combined branching fraction of 65\%.
They include modes with either one or three charged hadrons, and up to two neutral pions.
Neutral pions are reconstructed via their decay into pairs of photons detected in the ECAL.
The pattern of energy deposition in the ECAL typically occurs in ``strips", elongated in the $\phi$ direction as a result of interactions in the tracker material and the effect of the axial magnetic field.
The \tauh candidate is reconstructed from strips and charged hadrons, which are combined
using the mass ranges expected from the intermediate resonances.
A more detailed discussion of the HPS algorithm can be found
in~\cite{TAU-14-001}.
The reconstruction of hadronic tau decays has been optimized for tau leptons with large \pt where different tracks potentially merge.
This occurs because either the track reconstruction seed cannot be resolved or the tracks share so many hits that one track can not be reconstructed.
This leads to reconstructed decay modes with only two charged hadrons (instead of three) being accepted to accommodate the boosted topology.
The energy measurement of these high-\pt objects is dominated by the calorimeter and therefore has a good \pt resolution. The allowed mass range for the intermediate state reconstruction is broadened for
high-\pt tau leptons, to compensate for the mass resolution. With these adaptations the tau reconstruction efficiency is constant at $60\% \pm 6\%$ for $\pt>80$\GeV, as has been checked in simulations up to \pt = 3\TeV.
Hadronic tau decays identified by the HPS algorithm are required to be within the tracking acceptance, $\abs{\eta}<2.3$, and the tau \pt is required to be larger than 50\GeV to reduce the contamination from QCD jets.
Additionally the \pt of the  leading charged hadron is required to be larger than 20\GeV.
Subsequently, \tauh is distinguished from other objects
that could mimic
a tau candidate, such as QCD jets, electrons, or muons.
The discriminator against QCD jets is the most important, since the rate of QCD jets at the LHC is several orders of magnitude larger than the tau production rate.
Discrimination is based on isolation criteria: no additional PF charged hadrons or photons with $\abs{\sum \ptvec}$ above 2\GeV are allowed in an isolation cone of $\Delta R = \sqrt{\smash[b]{(\Delta \phi)^2
+(\Delta \eta)^2} }= 0.3$ (where $\phi$ is the azimuthal angle in radians and $\eta$ is the pseudorapidity)
around the \tauh candidate direction.
Particle-flow objects are corrected for additional collisions in the same bunch crossing (pileup).
Charged hadrons are identified as pileup objects by vertex association.
Neutral particle candidates are corrected by using an average \pt subtraction from the charged hadrons identified as pileup in a $\Delta R = 0.6$ cone. Details can be found in Ref.~\cite{TAU-14-001}.
Discrimination against electrons is obtained using a multivariate technique, based on various tau, photon, track and electron properties.
The muon discriminator searches for hits in the muon system associated with the track of the \tauh candidate. Both discriminators suppress light leptons by three orders of magnitude, without a significant reduction of the tau efficiency.
Events of interest for this analysis are required not to contain identified electrons or muons.
Electrons are required to satisfy shape and isolation criteria as well as $\pt>20\GeV$, and $\abs{\eta}<1.44$ or
$1.56<\abs{\eta}<2.50$. Muons  are required to be isolated and to have $\pt>20\GeV$ and $\abs{\eta}<2.4$.

\begin{figure}[htb]
\centering
\includegraphics[width=0.48\textwidth]{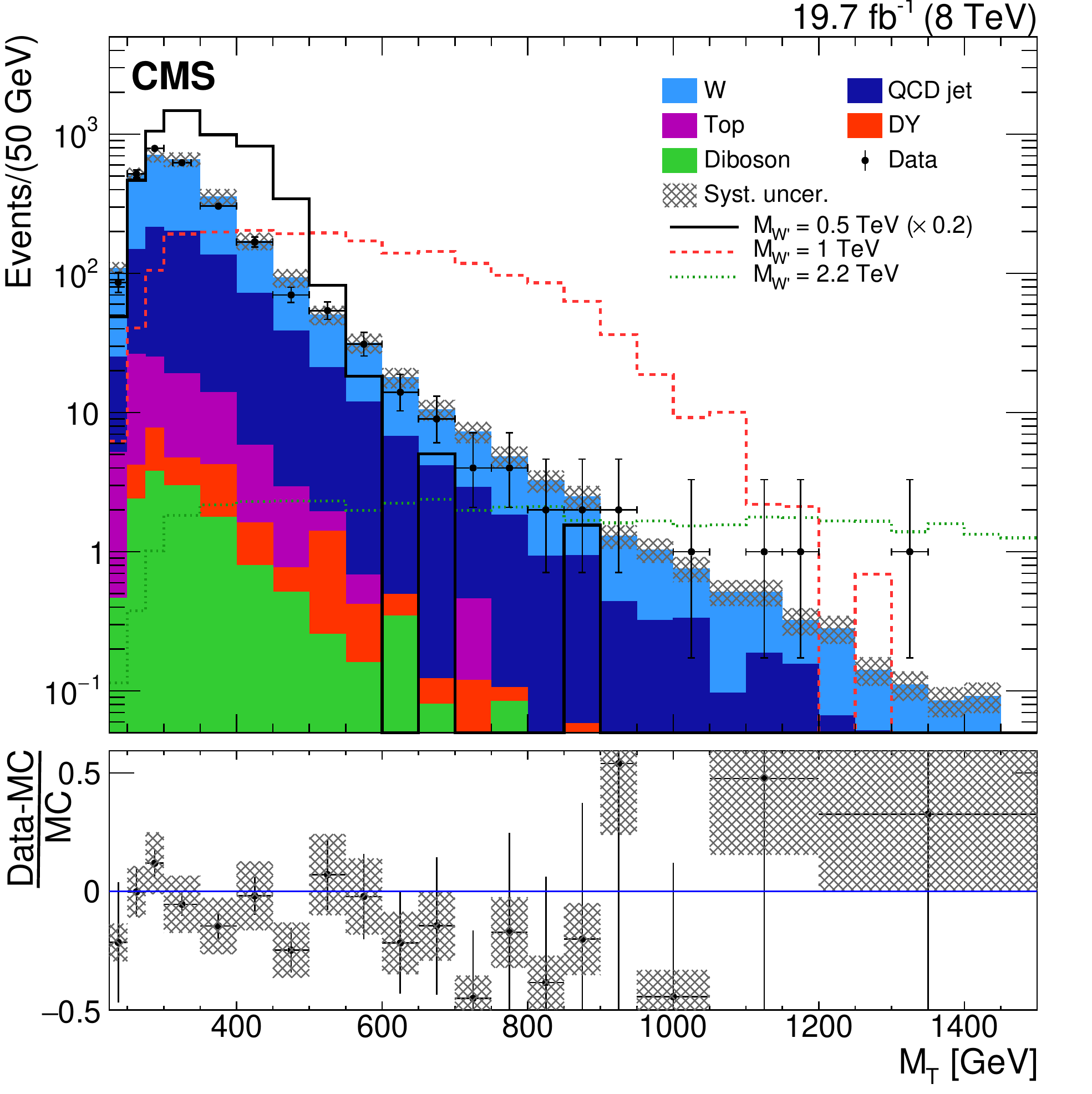}
\caption{\label{fig:finalmt}
The \MT distribution after the final selection. Data points with error bars show LHC data. The horizontal error bar
on each point indicates the width of the bin, which is 25\GeV for the first three bins and 50\GeV for all other
bins. The filled histogram shows the background estimate discussed in the text, and the hatched area the
uncertainty in this estimate.
The signal shapes for different SSM \PWpr boson masses are shown as open histograms. The cross section for SSM
$M_{\PWpr} = 500\GeV$ is scaled by 0.2.
In the ratio plot the bin-width is increased where needed to have at least one expected
background event in each
bin.
}
\end{figure}

\section{Analysis Strategy}
The strategy of this analysis is to select a heavy boson decaying almost at rest into \tauh and \MET.
In the tau channel, the impact of the interference between \PWpr and \PW bosons is expected to be substantially lower than that previously found in the electron and muon channels~\cite{EXO-12-060}. This occurs because the signal shape of a \PWpr boson with hadronically decaying tau leptons does not show a Jacobian peak structure,
because of the presence of two neutrinos
in the final state. The interference effect has therefore not been considered in this analysis.
For the ``jet+\MET" trigger, analysis thresholds of $\pt > 100\GeV$ for the leading jet and $\MET>140\GeV$ are
applied to account for differences of trigger and reconstructed energy definitions.
These analysis thresholds on the tau \pt and \MET, along with the kinematic selection on the ratio of $\pTtau/\MET$, yield an implicit lower threshold on the transverse mass.
The event is required to contain one isolated tau lepton.
Two kinematic criteria are applied to select signal events:
the ratio of the \tauh \pt to the \MET is required to satisfy $0.7 < \pTtau/\MET < 1.5$ and the angle
$\Delta \phi(\tau,\ptvecmiss)$ has to be greater than 2.4 radians.
This event selection mainly reduces the background in the low-\MT region, which has the largest background, while the signal efficiency at high \PWpr masses is only reduced by about 5\%.
The efficiency and acceptance for a \PWprTauNu event depend on the mass. For $M_{\PWpr} = 2.2\TeV$, 21\% of the events pass all identification and selection criteria. This reduces to
17\% for $M_{\PWpr} = 1\TeV$, 7\% for $M_{\PWpr} = 0.5\TeV$, and, at higher masses, to 16\% at $M_{\PWpr} = 3\TeV$.
The reduction for lower masses occurs because of the change in shape of the \MT distribution illustrated in Fig~\ref{fig:finalmt},
while for higher masses the off-shell production becomes dominant and shifts the events to lower \MT.
From the simulation of hadronic tau events with large \MT values,
above the kinematic turn-on, 42\% are accepted once all selection and identification criteria are taken into account.
This acceptance is independent of the \PWpr mass.
For the example case of \PWprTauNu with $M_{\PWpr} = 2.2\TeV$, the cross section calculated in the SSM is 13.5\unit{fb}. This yields 54.8 predicted signal events in the \tauh + \MET final state,
with the 21\% acceptance quoted above for this
$M_{\PWpr}$ value. The variation of the predicted SSM cross section with \PWpr mass can be seen in Fig.~\ref{fig:ssmlimit}.

\section{Background Estimation}
\label{sec:background}

The transverse mass distribution with the observed data and expected background events and uncertainties is
shown in Fig.~\ref{fig:finalmt} and Table~\ref{tab:lK8_Tab_BG2}.
The dominant background, contributing almost two thirds of the total, comes from the
off-shell tail of the SM \PW boson.
This background is indistinguishable from the signal, and is estimated from simulation.
The contribution from $\PW\to \Pe/\mu +\nu$ events, in which the electron or muon is not identified, is also taken from simulation.
The background contribution from events with one QCD jet falsely identified as a \tauh is suppressed by the $\pTtau/\MET$ requirement.
Nonetheless it is the second largest background for this search and is estimated from data using reference regions, separated from the signal region using the uncorrelated quantities, $\pTtau/\MET$ and \tauh isolation. The shape of the QCD jet background is estimated using
data events with a jet identified as a \tauh, fulfilling all kinematic criteria described earlier, apart from the isolation requirement. Its normalization is based on the ratio of the numbers of events with an isolated \tauh
($N_{\text{iso}}$) to those containing a non-isolated \tauh ($N_{\text{non-iso}}$), determined in a signal-free reference region with $\pTtau/\MET>1.5$.
This ratio is evaluated as a function of the hadronic decay modes of the tau lepton.
The mean ratio of isolated to non-isolated events is $R = N_{\text{iso}}/N_{\text{non-iso}} = 0.0066 \pm 0.12\%\stat\pm 0.16\%\syst$. Here the
contribution of non-QCD events is subtracted. It amounts to 24\% for $N_{\text{iso}}$ and 11\% for $N_{\text{non-iso}}$. The systematic uncertainty is estimated
by
changing the $\pTtau/\MET$ threshold and the variable in which the ratio $R$ is binned. The number of QCD jet events in the signal region is estimated, using this method, to be $620 \pm 124$ after
subtracting the contamination of 32\% from electroweak background events.
An additional systematic uncertainty of 20\% is included, derived from the normalization uncertainty in the electroweak background.
Other sources of background considered include top quark production, either in pairs or singly; Drell--Yan (DY) events; and tau leptons produced in diboson ($\PW\PW$, $\PW\PZ$, $\PZ\PZ$) events.
A large fraction of these are suppressed by requiring the back-to-back decay topology.
These backgrounds are shown in Fig.~\ref{fig:finalmt} as Top, DY, and Diboson, respectively. They contribute a total of 3\% to the background.
\begin{table*}[htb]
\centering
\topcaption{The event yields for observed data and estimated backgrounds, and the
product of acceptance and efficiency for the signal (\PWprTauNu)
for different threshold values \MTlower.}
\resizebox{\textwidth}{!}{
\begin{tabular}{r r . . . . . c .}
\multicolumn{1}{c}{$\MTlower [\GeVns{}]$}  & \multicolumn{1}{c}{Data} & \multicolumn{1}{c}{$VV$} &\multicolumn{1}{c}{DY} &\multicolumn{1}{c}{Top} &\multicolumn{1}{c}{QCD jets} &\multicolumn{1}{c}{\PW} & Sum of  & \multicolumn{1}{c}{Efficiency for}\\
& &&&&&&backgrounds&\multicolumn{1}{c}{$M_{\PWpr} = 2.2\TeV$}\\
\hline
200 &1990 &10 &10 &54 &620 &1380 &2080 $\pm$ 34\stat $\pm$ 250\syst & 0.21 \\
400 &364 &2.3 &2.9 &7.6 &151 &234 &398 $\pm$ 5.5\stat $\pm$ 63\syst & 0.19 \\
600 &41 &0.61 &0.37 &0.34 &18.2 &32.2 &51.7 $\pm$ 1.3\stat $\pm$ 9\syst & 0.16 \\
800 &10 &0.064 &0.072 &0 &3.6 &7.4 &11.1 $\pm$ 0.49\stat $\pm$ 2.1\syst & 0.12 \\
1000 &4 &0.0091 &0.027 &0 &1.07 &1.94 &3.05 $\pm$ 0.19\stat $\pm$ 0.66\syst & 0.096 \\
1200 &1 &0.0031 &0.016 &0 &0.31 &0.61 &0.94 $\pm$ 0.095\stat $\pm$ 0.22\syst & 0.071 \\
1400 &0 &0.0011 &0.0076 &0 &0.130 &0.180 &0.319 $\pm$ 0.046\stat $\pm$ 0.081\syst & 0.047 \\
\end{tabular}
\label{tab:lK8_Tab_BG2}
}
\end{table*}

\section{Systematic Uncertainties}
Most of the systematic uncertainties in this analysis affect the shape of the \MT distribution by changing the background and signal predictions.
Others influence the overall normalization; these include the uncertainty of 2.6\%~\cite{LUM-13-001} in the integrated luminosity.
Simulated event samples are used to evaluate shape-dependent uncertainties arising from the measurement of individual particles and jets in the events.
The kinematic variables of the individual objects are varied and the effect of the changes on the final \MT distribution is evaluated.
In the following, the shape-dependent uncertainties are listed in decreasing order of their importance for the high-\MT region.
The main uncertainty in the background yield for $\MT \geq 1$\TeV is due to the momentum measurement of the tau lepton~\cite{TauReco}, important for estimating the contribution from off-shell SM $\PW$-boson decays.
Using $\PZ\to \tau \tau$ events and
tau-mass fits, the uncertainty in the momentum scale is estimated to be 3\% of the tau \pt.
This estimation is confirmed by comparing energy measurements from tau and jet reconstruction algorithms for high-\pt taus. This results in a 15\% scale uncertainty in the background event yield, primarily from the tail of off-shell SM $\PW$ bosons, which is correlated with the uncertainty in the signal prediction.
There is an 8\% uncertainty in the event yield from the theoretical prediction of the background.
One contribution to this theory uncertainty comes from the NNLO QCD and NLO electroweak calculations and is evaluated following the prescription described in Ref.~\cite{leshouches2013SM}; there is an
additional contribution from the PDFs, for which the prescriptions of Refs.~\cite{Alekhin:2011sk,Botje:2011sn} are used.
The uncertainty in the event yield from the jet energy calibration is estimated to be 6\%. The calibration uncertainty is dependent on the jet $\eta$ and \pt, and is determined using dijet and
$\PZ\to \mu\mu +\text{jets}$ events~\cite{CMS-JME-10-011}.
The knowledge of the reconstruction efficiency for high-\pt tau leptons is a source of uncertainty influencing the background and signal
normalization. The efficiency is determined by studying $\PZ\to \tau \tau$ and $\ttbar$ processes~\cite{TauReco}. The resulting uncertainty in the normalization is 6\%.
There is an uncertainty of 20\% in the QCD jet contribution to the background, which is estimated from statistical uncertainties in the control regions and cross checks of the method, and which results in a 4--6\% uncertainty in the overall background yield.
Other sources of uncertainty are the jet energy resolution ($\eta$ and \pt dependent)~\cite{CMS-JME-10-011}, pileup modeling (5\% on the estimated number of additional interactions), and other factors affecting the \MET determination, such as low-energy deposits not associated with a jet (10\% uncertainty in the energy of deposits smaller than 10\GeV). The overall impact of these effects is a 6\% background uncertainty.
The impact of all these uncertainties on the signal acceptance has been evaluated using the simulated samples.
The size and relative importance of the effects observed are similar to those for the
background yield, and depend on the shape of the \MT distribution.

\section{Results}
\label{sec:limit}

The final transverse mass distribution in Fig.~\ref{fig:finalmt} shows no significant deviations from the predicted background.
A multibin approach is used to derive a limit on the \PWpr-boson mass. A likelihood function is evaluated separately using the numbers of events in each \MT bin. The likelihood functions from all bins are
combined to extract the mass limit.
For a more model-independent limit, a single-bin approach is used, counting all events above a threshold \MTlower and comparing the number with the expected SM background.
The parameter of interest is the product of the signal cross section and the branching fraction, $\sigma \, \mathcal{B}(\PWprTauNu)$. Limits are obtained at 95\% confidence level (\CL) using a Bayesian approach~\cite{BayesianLimit} with a
uniform prior.

\begin{figure}[!htb]
\centering
\includegraphics[width=0.48\textwidth]{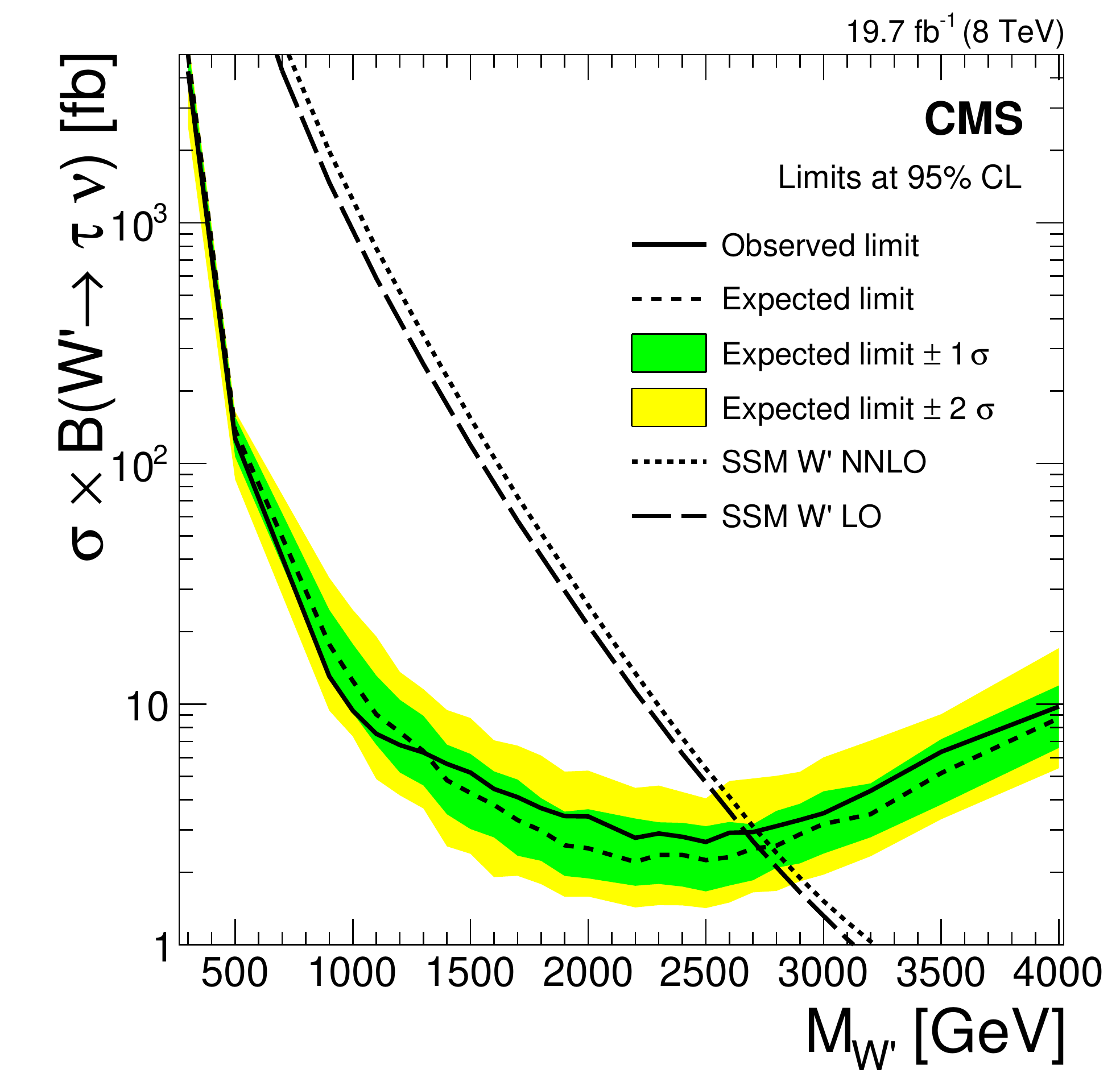}
\caption{
Limits on the product of cross section and branching fraction into $\tau \nu$ for a SSM \PWpr boson. The solid line
shows the limit observed with 19.7\fbinv of data while
the dashed line corresponds to the expected limit.
The shaded bands indicate the 68\% and 95\% confidence intervals of the expected limit.
The dotted and the long-dashed lines show the cross section prediction in the SSM as a function of the \PWpr boson
mass, in NNLO and LO, respectively.
}
\label{fig:ssmlimit}
\end{figure}

\begin{figure}[!htb]
\centering
\includegraphics[width=0.48\textwidth]{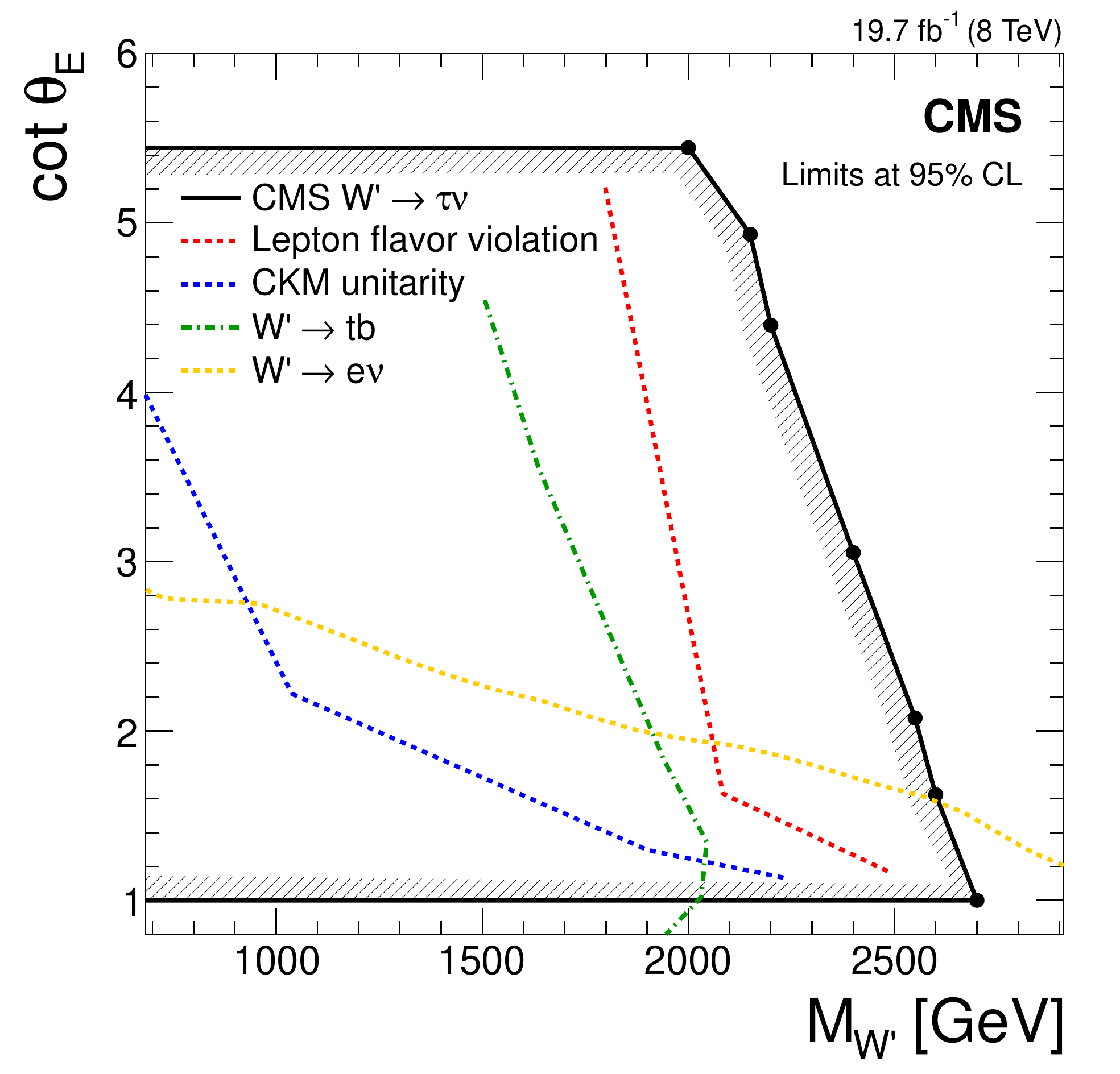}
\caption{
Limits on the NUGIM parameter space are shown from various analyses. The solid line refers to this analysis.
The non-LHC limits (CKM and Lepton flavor violation) are calculated in Ref.~\cite{Kim:2011qk}. The \PWpr results are from Ref.~\cite{Chatrchyan:2014koa}
for the tb final state and Ref.~\cite{EXO-12-060} for $\Pe\nu$ as
reinterpreted in Ref.~\cite{AlexLisaPaper}. The lines correspond to 95\% \CL limits.
}
\label{fig:nugimlimit}
\end{figure}

The limit on $\sigma \, \mathcal{B}(\PWprTauNu)$ as a function of the SSM \PWpr-boson mass is shown in Fig.~\ref{fig:ssmlimit}. The observed and expected limits are in agreement.
The SSM \PWpr boson is excluded for masses $0.3< M_{\PWpr} < 2.7\TeV$ at 95\% \CL in the tau channel. The lower mass limit is due to the trigger threshold and rising background.
The \PWpr mass limit obtained at 95\% \CL is 400\GeV lower for a signal cross section calculated to leading order.
In the high mass region, off-shell production of \PWpr bosons becomes dominant, shifting the signal \MT distribution to lower \MT. In comparison, analyses of the muon and electron channels have set limits of 3.0 and
3.2\TeV on the SSM \PWpr mass, respectively~\cite{EXO-12-060}.
In addition to the limit on the SSM \PWpr boson, limits are set on the parameter space of the NUGIM. Only leading-order signal cross sections are available in the NUGIM. A separate cross section limit is derived for each value of the
model parameter $\cot \theta_E$, since the signal efficiency depends on this parameter. The actual width of the \PWpr resonance for a given mass, as shown in Fig.~\ref{fig:BRPlot}, is taken into account.
From these limits, constraints on the mass of the \PWpr boson as a function of the coupling parameter $\cot \theta_E$ are derived in the same way as described previously for the SSM \PWpr boson.
The resulting constraints from these mass exclusion limits on the parameter space can be seen in Fig.~\ref{fig:nugimlimit}.
The \PWpr mass limit is 2.0\TeV for $\cot \theta_E = 5.5$, rising to a \PWpr boson mass of 2.7\TeV for $\cot \theta_E = 1$.
This variation is due in part to the change in coupling strength to the tau lepton, which affects the decay, as shown in Fig.~\ref{fig:BRPlot},
and in part to the change in coupling to light quarks, which affects the production.
For $\cot \theta_E > 5.5$ the width of the \PWpr becomes very broad, and large virtual corrections are needed. This search sets significantly better limits than the previous constraints from
direct and indirect searches for large $\cot \theta_E$~\cite{Kim:2011qk,Chatrchyan:2014koa,EXO-12-060} reinterpreted in~\cite{AlexLisaPaper}. For $\cot \theta_E < 1$, the light families yield a better
sensitivity because of their higher efficiency and branching fraction as shown for the case of the electron channel in Fig.~\ref{fig:nugimlimit}.

\begin{figure}[htb]
\centering
\includegraphics[width=0.48\textwidth]{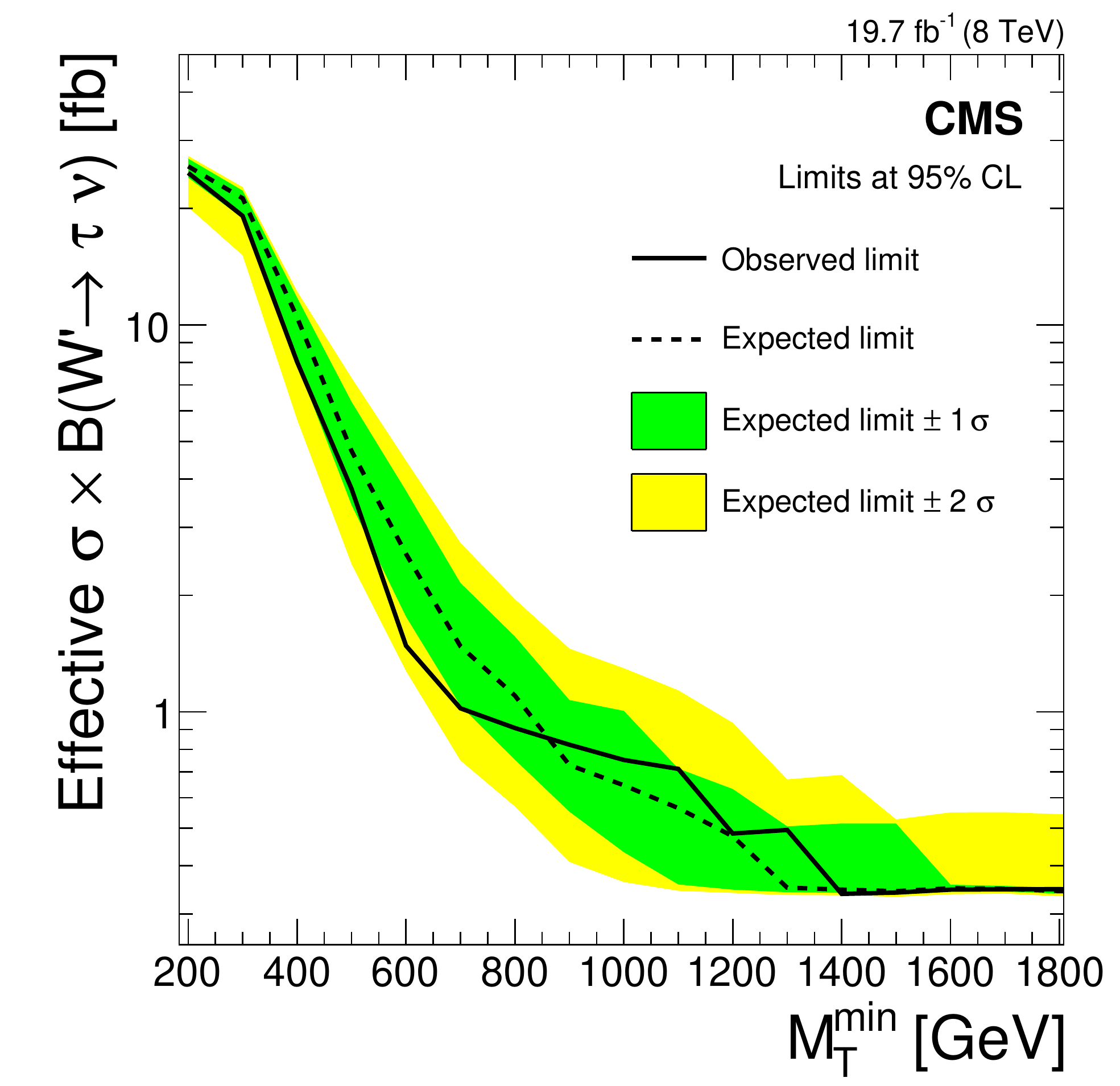}
\caption{\label{fig:MILimit} Model independent limits, on the effective cross section for a \PWpr-like signal above a threshold value \MTlower, for different \MTlower.
The solid line shows the limit observed with 19.7\fbinv of data while the dashed line corresponds to the expected limit.
The shaded bands indicate the 68\% and 95\% confidence intervals of the expected limit.
The region above the curve is excluded.
}
\end{figure}

The multibin approach assumes a certain signal shape in \MT. However, new physics processes yielding a tau+\MET final state could cause an excess of a different shape. To be independent of models, a single-bin approach compares the number of observed events above a sliding \MT threshold, denoted \MTlower, with the SM expectation for this \MT range. The resulting cross section limit as a function of \MTlower
is shown in Fig.~\ref{fig:MILimit}.
The reconstruction efficiency is estimated to be 42\% for \PWpr events satisfying the condition $\MT > \MTlower$.
It may be noted that the fraction of the signal that satisfies the \MTlower requirement depends on the particular model, and
is mass-dependent.
The reconstruction efficiency has an uncertainty corresponding to that of a typical \PWpr-like signal at different \MTlower thresholds.
This allows a reinterpretation in various models by evaluating
the signal efficiency, $\varepsilon_{\text{signal}}$, for the \MTlower threshold, defined as the number of events in the signal region with $\MT>\MTlower$ divided by the total number of generated events:
$\varepsilon_{\text{signal}} = N_{\MT>\MTlower}/N_\text{total}$.

\section{Summary}

In summary, the first search for an excess in the transverse mass distribution of the tau+\MET channel has been performed. The data sample was collected with the
CMS detector in proton-proton
collisions at $\sqrt{s} = 8\TeV$, and corresponds to an integrated luminosity of 19.7\fbinv. No significant excess beyond the SM expectation is observed. An SSM \PWpr
boson is excluded in the mass range $0.3\TeV  < M_{\PWpr} < 2.7$\TeV
at 95\% confidence level. Within the NUGIM the lower limit on the \PWpr-boson mass depends on the coupling constant $\cot \theta_E$ and varies from 2.0 to 2.7\TeV at
95\% confidence level.

\begin{acknowledgments}
We congratulate our colleagues in the CERN accelerator departments for the excellent performance of the LHC and thank the technical and administrative staffs at CERN and at other CMS institutes for their contributions to the success of the CMS effort. In addition, we gratefully acknowledge the computing centers and personnel of the Worldwide LHC Computing Grid for delivering so effectively the computing infrastructure essential to our analyses. Finally, we acknowledge the enduring support for the construction and operation of the LHC and the CMS detector provided by the following funding agencies: BMWFW and FWF (Austria); FNRS and FWO (Belgium); CNPq, CAPES, FAPERJ, and FAPESP (Brazil); MES (Bulgaria); CERN; CAS, MoST, and NSFC (China); COLCIENCIAS (Colombia); MSES and CSF (Croatia); RPF (Cyprus); MoER, ERC IUT and ERDF (Estonia); Academy of Finland, MEC, and HIP (Finland); CEA and CNRS/IN2P3 (France); BMBF, DFG, and HGF (Germany); GSRT (Greece); OTKA and NIH (Hungary); DAE and DST (India); IPM (Iran); SFI (Ireland); INFN (Italy); MSIP and NRF (Republic of Korea); LAS (Lithuania); MOE and UM (Malaysia); CINVESTAV, CONACYT, SEP, and UASLP-FAI (Mexico); MBIE (New Zealand); PAEC (Pakistan); MSHE and NSC (Poland); FCT (Portugal); JINR (Dubna); MON, RosAtom, RAS and RFBR (Russia); MESTD (Serbia); SEIDI and CPAN (Spain); Swiss Funding Agencies (Switzerland); MST (Taipei); ThEPCenter, IPST, STAR and NSTDA (Thailand); TUBITAK and TAEK (Turkey); NASU and SFFR (Ukraine); STFC (United Kingdom); DOE and NSF (USA).
\end{acknowledgments}

\bibliography{auto_generated}

\providecommand{\href}[2]{#2}\begingroup\raggedright\begin{thebibliography}{10}%
\makeatletter
\providecommand{\hrefCMSnoop }[0]{\@secondoftwo}%
\makeatother
\providecommand{\doi}{\texttt{doi:}\begingroup \urlstyle{tt}\Url}

\bibitem{reference-model}
\hrefCMSnoop {}{G.~Altarelli, B.~Mele, and M.~Ruiz-Altaba, ``Searching for new
  heavy vector bosons in $\mathrm{p\bar{p}}$ colliders'',} \textit{ Z. Phys. C}
  \textbf{ 45} (1989) 109,
  \href{http://dx.doi.org/10.1007/BF01556677}{\doi{10.1007/BF01556677}}.

\bibitem{Chiang:2009kb}
\hrefCMSnoop {}{C.-W. Chiang, N.~G. Deshpande, X.-G. He, and J.~Jiang, ``Family
  {$SU(2)_l \times SU(2)_h \times U(1)$} model'',} \textit{ Phys. Rev. D}
  \textbf{ 81} (2010) 015006,
  \href{http://dx.doi.org/10.1103/PhysRevD.81.015006}{\doi{10.1103/PhysRevD.81.015006}},
\href{http://www.arXiv.org/abs/0911.1480}{\texttt{arXiv:0911.1480}}.
%%CITATION = ARXIV:0911.1480;%%.

\bibitem{Li:1981nk}
\hrefCMSnoop {}{X.~Li and E.~Ma, ``{Gauge Model of Generation
  Nonuniversality}'',} \textit{ Phys. Rev. Lett.} \textbf{ 47} (1981) 1788,
\href{http://dx.doi.org/10.1103/PhysRevLett.47.1788}{\doi{10.1103/PhysRevLett.47.1788}}.
%%CITATION = PRLTA,47,1788;%%.

\bibitem{Muller:1996dj}
\hrefCMSnoop {}{D.~J. Muller and S.~Nandi, ``{Top flavor: A separate SU(2) for
  the third family}'',} \textit{ Phys. Lett. B} \textbf{ 383} (1996) 345,
  \href{http://dx.doi.org/10.1016/0370-2693(96)00745-9}{\doi{10.1016/0370-2693(96)00745-9}},
\href{http://www.arXiv.org/abs/hep-ph/9602390}{\texttt{arXiv:hep-ph/9602390}}.
%%CITATION = HEP-PH/9602390;%%.

\bibitem{Malkawi:1996fs}
\hrefCMSnoop {}{E.~Malkawi, T.~M.~P. Tait, and C.~P. Yuan, ``A model of strong
  flavor dynamics for the top quark'',} \textit{ Phys. Lett. B} \textbf{ 385}
  (1996) 304,
  \href{http://dx.doi.org/10.1016/0370-2693(96)00859-3}{\doi{10.1016/0370-2693(96)00859-3}},
\href{http://www.arXiv.org/abs/hep-ph/9603349}{\texttt{arXiv:hep-ph/9603349}}.
%%CITATION = HEP-PH/9603349;%%.

\bibitem{cdf-limit}
\hrefCMSnoop {}{{CDF} Collaboration, ``{Search for a new heavy gauge boson $W'$
  with even signature electron + missing missing transverse energy in
  $\mathrm{p \bar{p}}$ collisions at $\sqrt{s} = 1.96$ TeV}'',} \textit{ Phys.
  Rev. D} \textbf{ 83} (2011) 031102,
  \href{http://dx.doi.org/10.1103/PhysRevD.83.031102}{\doi{10.1103/PhysRevD.83.031102}},
\href{http://www.arXiv.org/abs/1012.5145}{\texttt{arXiv:1012.5145}}.
%%CITATION = ARXIV:1012.5145;%%.

\bibitem{d0-limit}
\hrefCMSnoop {}{{D0} Collaboration, ``Search for {$W^\prime$} Bosons Decaying
  to an Electron and a Neutrino with the {D0} Detector'',} \textit{ Phys. Rev.
  Lett.} \textbf{ 100} (2008) 031804,
  \href{http://dx.doi.org/10.1103/PhysRevLett.100.031804}{\doi{10.1103/PhysRevLett.100.031804}},
  \href{http://www.arXiv.org/abs/0710.2966}{\texttt{arXiv:0710.2966}}.

\bibitem{ATLAS-paper-14}
\hrefCMSnoop {}{{ATLAS Collaboration}, ``{Search for new particles in events
  with one lepton and missing transverse momentum in pp collisions at
  $\sqrt{s}$ = 8 TeV with the ATLAS detector}'',} \textit{ JHEP} \textbf{ 09}
  (2014) 37,
  \href{http://dx.doi.org/10.1007/JHEP09(2014)037}{\doi{10.1007/JHEP09(2014)037}},
\href{http://www.arXiv.org/abs/1407.7494}{\texttt{arXiv:1407.7494}}.
%%CITATION = ARXIV:1407.7494;%%.

\bibitem{EXO-12-060}
\hrefCMSnoop {}{{CMS Collaboration}, ``{Search for physics beyond the standard
  model in final states with a lepton and missing transverse energy in
  proton-proton collisions at $\sqrt{s} = 8$ TeV}'',} (2014).
  \href{http://www.arXiv.org/abs/1408.2745}{\texttt{arXiv:1408.2745}}.
Submitted to Phys. Rev. D.
%%CITATION = ARXIV:1408.2745;%%.

\bibitem{cms}
\hrefCMSnoop {}{{CMS Collaboration}, ``The {CMS} experiment at the {CERN}
  {LHC}'',} \textit{ JINST} \textbf{ 3} (2008) S08004,
\href{http://dx.doi.org/10.1088/1748-0221/3/08/S08004}{\doi{10.1088/1748-0221/3/08/S08004}}.
%%CITATION = JINST,3,S08004;%%.

\bibitem{Kim:2011qk}
\hrefCMSnoop {}{Y.~G. Kim and K.~Y. Lee, ``{Early LHC bound on $\mathrm{W}'$
  boson in the nonuniversal gauge interaction model}'',} \textit{ Phys. Lett.
  B} \textbf{ 706} (2012) 367,
  \href{http://dx.doi.org/10.1016/j.physletb.2011.11.032}{\doi{10.1016/j.physletb.2011.11.032}},
\href{http://www.arXiv.org/abs/1105.2653}{\texttt{arXiv:1105.2653}}.
%%CITATION = ARXIV:1105.2653;%%.

\bibitem{AlexLisaPaper}
\hrefCMSnoop {}{L.~Edelh{\"a}user and A.~Knochel, ``{Observing nonstandard
  W$^\prime$ and Z$^\prime$ through the third generation and Higgs lens}'',}
  (2014).
\href{http://www.arXiv.org/abs/1408.0914}{\texttt{arXiv:1408.0914}}.
%%CITATION = ARXIV:1408.0914;%%.

\bibitem{Chatrchyan:2014koa}
\hrefCMSnoop {}{{CMS Collaboration}, ``{Search for $\mathrm{W}'\to \mathrm{tb}$
  decays in the lepton + jets final state in pp collisions at $\sqrt{s} = 8$
  TeV}'',} \textit{ JHEP} \textbf{ 05} (2014) 108,
  \href{http://dx.doi.org/10.1007/JHEP05(2014)108}{\doi{10.1007/JHEP05(2014)108}},
\href{http://www.arXiv.org/abs/1402.2176}{\texttt{arXiv:1402.2176}}.
%%CITATION = ARXIV:1402.2176;%%.

\bibitem{dudkov}
\hrefCMSnoop {}{E.~Boos, V.~Bunichev, L.~Dudko, and M.~Perfilov,
  ``{Interference between W$^\prime$ and W in single-top quark production
  processes}'',} \textit{ Phys. Lett. B} \textbf{ 655} (2007) 245,
  \href{http://dx.doi.org/10.1016/j.physletb.2007.03.064}{\doi{10.1016/j.physletb.2007.03.064}},
\href{http://www.arXiv.org/abs/hep-ph/0610080}{\texttt{arXiv:hep-ph/0610080}}.
%%CITATION = HEP-PH/0610080;%%.

\bibitem{criticalPaper}
E.~Accomando\hrefCMSnoop {}{ {et~al.}, ``{Interference effects in heavy
  $W'$-boson searches at the LHC}'',} \textit{ Phys. Rev. D} \textbf{ 85}
  (2012) 115017,
  \href{http://dx.doi.org/10.1103/PhysRevD.85.115017}{\doi{10.1103/PhysRevD.85.115017}},
\href{http://www.arXiv.org/abs/1110.0713}{\texttt{arXiv:1110.0713}}.
%%CITATION = ARXIV:1110.0713;%%.

\bibitem{WprimeHelicity}
\hrefCMSnoop {}{T.~G. Rizzo, ``The determination of the helicity of
  {W$^\prime$} boson couplings at the {LHC}'',} \textit{ JHEP} \textbf{ 05}
  (2007) 037,
  \href{http://dx.doi.org/10.1088/1126-6708/2007/05/037}{\doi{10.1088/1126-6708/2007/05/037}},
  \href{http://www.arXiv.org/abs/0704.0235}{\texttt{arXiv:0704.0235}}.

\bibitem{madgraph}
J.~Alwall\hrefCMSnoop {}{ {et~al.}, ``{MadGraph 5: going beyond}'',} \textit{
  JHEP} \textbf{ 06} (2011) 128,
  \href{http://dx.doi.org/10.1007/JHEP06(2011)128}{\doi{10.1007/JHEP06(2011)128}},
\href{http://www.arXiv.org/abs/1106.0522}{\texttt{arXiv:1106.0522}}.
%%CITATION = ARXIV:1106.0522;%%.

\bibitem{Sjostrand:2006za}
\hrefCMSnoop {}{T.~Sj{\"o}strand, S.~Mrenna, and P.~Z. Skands, ``{PYTHIA} 6.4
  physics and manual'',} \textit{ JHEP} \textbf{ 05} (2006) 26,
  \href{http://dx.doi.org/10.1088/1126-6708/2006/05/026}{\doi{10.1088/1126-6708/2006/05/026}},
\href{http://www.arXiv.org/abs/hep-ph/0603175}{\texttt{arXiv:hep-ph/0603175}}.
%%CITATION = HEP-PH/0603175;%%.

\bibitem{Nason:2004rx}
\hrefCMSnoop {}{P.~Nason, ``{A new method for combining NLO QCD with shower
  Monte Carlo algorithms}'',} \textit{ JHEP} \textbf{ 11} (2004) 040,
  \href{http://dx.doi.org/10.1088/1126-6708/2004/11/040}{\doi{10.1088/1126-6708/2004/11/040}},
\href{http://www.arXiv.org/abs/hep-ph/0409146}{\texttt{arXiv:hep-ph/0409146}}.
%%CITATION = HEP-PH/0409146;%%.

\bibitem{Frixione:2007vw}
\hrefCMSnoop {}{S.~Frixione, P.~Nason, and C.~Oleari, ``{Matching NLO QCD
  computations with Parton Shower simulations: the POWHEG method}'',} \textit{
  JHEP} \textbf{ 11} (2007) 070,
  \href{http://dx.doi.org/10.1088/1126-6708/2007/11/070}{\doi{10.1088/1126-6708/2007/11/070}},
\href{http://www.arXiv.org/abs/0709.2092}{\texttt{arXiv:0709.2092}}.
%%CITATION = 0709.2092;%%.

\bibitem{Re:2010bp}
\hrefCMSnoop {}{E.~Re, ``{Single-top $W$ $t$-channel production matched with
  parton showers using the POWHEG method}'',} \textit{ Eur. Phys. J. C}
  \textbf{ 71} (2011) 1547,
  \href{http://dx.doi.org/10.1140/epjc/s10052-011-1547-z}{\doi{10.1140/epjc/s10052-011-1547-z}},
\href{http://www.arXiv.org/abs/1009.2450}{\texttt{arXiv:1009.2450}}.
%%CITATION = ARXIV:1009.2450;%%.

\bibitem{Frederix:2012dh}
\hrefCMSnoop {}{R.~Frederix, E.~Re, and P.~Torrielli, ``{Single-top $t$-channel
  hadroproduction in the four-flavour scheme with POWHEG and aMC@NLO}'',}
  \textit{ JHEP} \textbf{ 09} (2012) 130,
  \href{http://dx.doi.org/10.1007/JHEP09(2012)130}{\doi{10.1007/JHEP09(2012)130}},
\href{http://www.arXiv.org/abs/1207.5391}{\texttt{arXiv:1207.5391}}.
%%CITATION = ARXIV:1207.5391;%%.

\bibitem{Alioli:2011as}
\hrefCMSnoop {}{S.~Alioli, S.-O. Moch, and P.~Uwer, ``{Hadronic top-quark
  pair-production with one jet and parton showering}'',} \textit{ JHEP}
  \textbf{ 01} (2012) 137,
  \href{http://dx.doi.org/10.1007/JHEP01(2012)137}{\doi{10.1007/JHEP01(2012)137}},
\href{http://www.arXiv.org/abs/1110.5251}{\texttt{arXiv:1110.5251}}.
%%CITATION = ARXIV:1110.5251;%%.

\bibitem{tauola}
N.~Davidson\hrefCMSnoop {}{ {et~al.}, ``Universal interface of {TAUOLA}:
  Technical and physics documentation'',} \textit{ Comput. Phys. Commun.}
  \textbf{ 183} (2012) 821,
  \href{http://dx.doi.org/10.1016/j.cpc.2011.12.009}{\doi{10.1016/j.cpc.2011.12.009}},
\href{http://www.arXiv.org/abs/1002.0543}{\texttt{arXiv:1002.0543}}.
%%CITATION = ARXIV:1002.0543;%%.

\bibitem{geant4}
\hrefCMSnoop {}{{GEANT4} Collaboration, ``{GEANT4---a simulation toolkit}'',}
  \textit{ Nucl. Instrum. Meth. A} \textbf{ 506} (2003) 250,
\href{http://dx.doi.org/10.1016/S0168-9002(03)01368-8}{\doi{10.1016/S0168-9002(03)01368-8}}.
%%CITATION = NUIMA,A506,250;%%.

\bibitem{mcsanc}
\hrefCMSnoop {}{S.~G. Bondarenko and A.~A. Sapronov, ``{NLO EW and QCD
  proton-proton cross section calculations with mcsanc-v1.01}'',} \textit{
  Comput. Phys. Commun.} \textbf{ 184} (2013) 2343,
  \href{http://dx.doi.org/10.1016/j.cpc.2013.05.010}{\doi{10.1016/j.cpc.2013.05.010}},
\href{http://www.arXiv.org/abs/1301.3687}{\texttt{arXiv:1301.3687}}.
%%CITATION = ARXIV:1301.3687;%%.

\bibitem{fewz}
\hrefCMSnoop {}{R.~Gavin, Y.~Li, F.~Petriello, and S.~Quackenbush, ``{W Physics
  at the LHC with FEWZ 2.1}'',} \textit{ Comput. Phys. Commun.} \textbf{ 184}
  (2013) 208,
  \href{http://dx.doi.org/10.1016/j.cpc.2012.09.005}{\doi{10.1016/j.cpc.2012.09.005}},
\href{http://www.arXiv.org/abs/1201.5896}{\texttt{arXiv:1201.5896}}.
%%CITATION = ARXIV:1201.5896;%%.

\bibitem{leshouches2013SM}
\hrefCMSnoop {}{J.~Butterworth {et~al.}, ``{Les Houches 2013: Physics at TeV
  Colliders: Standard Model Working Group Report}'',} (2014).
\href{http://www.arXiv.org/abs/1405.1067}{\texttt{arXiv:1405.1067}}.
%%CITATION = ARXIV:1405.1067;%%.

\bibitem{Czakon:2013goa}
\hrefCMSnoop {}{M.~Czakon, P.~Fiedler, and A.~Mitov, ``{Total Top-Quark
  Pair-Production Cross Section at Hadron Colliders Through
  $O(\alpha^{4}_{s})$}'',} \textit{ Phys. Rev. Lett.} \textbf{ 110} (2013)
  252004,
  \href{http://dx.doi.org/10.1103/PhysRevLett.110.252004}{\doi{10.1103/PhysRevLett.110.252004}},
\href{http://www.arXiv.org/abs/1303.6254}{\texttt{arXiv:1303.6254}}.
%%CITATION = ARXIV:1303.6254;%%.

\bibitem{Campbell:2011bn}
\hrefCMSnoop {}{J.~M. Campbell, R.~K. Ellis, and C.~Williams, ``{Vector boson
  pair production at the LHC}'',} \textit{ JHEP} \textbf{ 07} (2011) 18,
  \href{http://dx.doi.org/10.1007/JHEP07(2011)018}{\doi{10.1007/JHEP07(2011)018}},
\href{http://www.arXiv.org/abs/1105.0020}{\texttt{arXiv:1105.0020}}.
%%CITATION = ARXIV:1105.0020;%%.

\bibitem{Pumplin:2002vw}
J.~Pumplin\hrefCMSnoop {}{ {et~al.}, ``New generation of parton distributions
  with uncertainties from global {QCD} analysis'',} \textit{ JHEP} \textbf{ 07}
  (2002) 012,
  \href{http://dx.doi.org/10.1088/1126-6708/2002/07/012}{\doi{10.1088/1126-6708/2002/07/012}},
\href{http://www.arXiv.org/abs/hep-ph/0201195}{\texttt{arXiv:hep-ph/0201195}}.
%%CITATION = HEP-PH/0201195;%%.

\bibitem{Gao:2013xoa}
J.~Gao\hrefCMSnoop {}{ {et~al.}, ``{CT10 next-to-next-to-leading order global
  analysis of QCD}'',} \textit{ Phys. Rev. D} \textbf{ 89} (2014) 033009,
  \href{http://dx.doi.org/10.1103/PhysRevD.89.033009}{\doi{10.1103/PhysRevD.89.033009}},
\href{http://www.arXiv.org/abs/1302.6246}{\texttt{arXiv:1302.6246}}.
%%CITATION = ARXIV:1302.6246;%%.

\bibitem{Ball:2013hta}
\hrefCMSnoop {}{{NNPDF} Collaboration, ``{Parton distributions with QED
  corrections}'',} \textit{ Nucl. Phys. B} \textbf{ 877} (2013) 290,
  \href{http://dx.doi.org/10.1016/j.nuclphysb.2013.10.010}{\doi{10.1016/j.nuclphysb.2013.10.010}},
\href{http://www.arXiv.org/abs/1308.0598}{\texttt{arXiv:1308.0598}}.
%%CITATION = ARXIV:1308.0598;%%.

\bibitem{CMS-PAS-PFT-10-001}
\href {http://cdsweb.cern.ch/record/1247373}{{CMS Collaboration},
  ``Commissioning of the Particle-flow Event Reconstruction with the first
  {LHC} collisions recorded in the {CMS} detector'',} CMS Physics Analysis
  Summary CMS-PAS-PFT-10-001, 2010.

\bibitem{TAU-14-001}
\hrefCMSnoop {}{{CMS Collaboration}, ``Reconstruction and identification of tau
  lepton decays to hadrons and tau neutrino at {CMS}'',} (2015).
  \href{http://www.arXiv.org/abs/1510.07488}{\texttt{arXiv:1510.07488}}.
  submitted to JINST.

\bibitem{LUM-13-001}
\href {http://cdsweb.cern.ch/record/1598864}{{CMS Collaboration}, ``CMS
  Luminosity Based on Pixel Cluster Counting - Summer 2013 Update'',} CMS
  Physics Analysis Summary CMS-PAS-LUM-13-001, 2013.

\bibitem{TauReco}
\hrefCMSnoop {}{{CMS Collaboration}, ``{Performance of tau-lepton
  reconstruction and identification in CMS}'',} \textit{ JINST} \textbf{ 7}
  (2012) P01001,
  \href{http://dx.doi.org/10.1088/1748-0221/7/01/P01001}{\doi{10.1088/1748-0221/7/01/P01001}},
\href{http://www.arXiv.org/abs/1109.6034}{\texttt{arXiv:1109.6034}}.
%%CITATION = ARXIV:1109.6034;%%.

\bibitem{Alekhin:2011sk}
\hrefCMSnoop {}{S.~Alekhin {et~al.}, ``{The PDF4LHC Working Group Interim
  Report}'',} (2011).
\href{http://www.arXiv.org/abs/1101.0536}{\texttt{arXiv:1101.0536}}.
%%CITATION = ARXIV:1101.0536;%%.

\bibitem{Botje:2011sn}
M.~Botje\hrefCMSnoop {}{ {et~al.}, ``{The PDF4LHC Working Group Interim
  Recommendations}'',} (2011).
\href{http://www.arXiv.org/abs/1101.0538}{\texttt{arXiv:1101.0538}}.
%%CITATION = ARXIV:1101.0538;%%.

\bibitem{CMS-JME-10-011}
\hrefCMSnoop {}{{CMS Collaboration}, ``Determination of jet energy calibration
  and transverse momentum resolution in {CMS}'',} \textit{ JINST} \textbf{ 6}
  (2011) P11002,
  \href{http://dx.doi.org/10.1088/1748-0221/6/11/P11002}{\doi{10.1088/1748-0221/6/11/P11002}},
\href{http://www.arXiv.org/abs/1107.4277}{\texttt{arXiv:1107.4277}}.
%%CITATION = ARXIV:1107.4277;%%.

\bibitem{BayesianLimit}
\hrefCMSnoop {}{{Particle Data Group}, K.~A. Olive {et~al.}, ``{Review of
  Particle Physics}'',} \textit{ Chin. Phys. C} \textbf{ 38} (2014) 090001,
  \href{http://dx.doi.org/10.1088/1674-1137/38/9/090001}{\doi{10.1088/1674-1137/38/9/090001}}.
[Chap. 38].
%%CITATION = CHPHD,C38,090001;%%.

\end{thebibliography}\endgroup

\cleardoublepage \appendix\section{The CMS Collaboration \label{app:collab}}\begin{sloppypar}\hyphenpenalty=5000\widowpenalty=500\clubpenalty=5000\textbf{Yerevan Physics Institute,  Yerevan,  Armenia}\\*[0pt]
V.~Khachatryan, A.M.~Sirunyan, A.~Tumasyan
\vskip\cmsinstskip
\textbf{Institut f\"{u}r Hochenergiephysik der OeAW,  Wien,  Austria}\\*[0pt]
W.~Adam, E.~Asilar, T.~Bergauer, J.~Brandstetter, E.~Brondolin, M.~Dragicevic, J.~Er\"{o}, M.~Flechl, M.~Friedl, R.~Fr\"{u}hwirth\cmsAuthorMark{1}, V.M.~Ghete, C.~Hartl, N.~H\"{o}rmann, J.~Hrubec, M.~Jeitler\cmsAuthorMark{1}, V.~Kn\"{u}nz, A.~K\"{o}nig, M.~Krammer\cmsAuthorMark{1}, I.~Kr\"{a}tschmer, D.~Liko, T.~Matsushita, I.~Mikulec, D.~Rabady\cmsAuthorMark{2}, B.~Rahbaran, H.~Rohringer, J.~Schieck\cmsAuthorMark{1}, R.~Sch\"{o}fbeck, J.~Strauss, W.~Treberer-Treberspurg, W.~Waltenberger, C.-E.~Wulz\cmsAuthorMark{1}
% ...
\vskip\cmsinstskip
% ...
1:~~Also at Vienna University of Technology, Vienna, Austria\\
2:~~Also at CERN, European Organization for Nuclear Research, Geneva, Switzerland\\
% ...

\end{sloppypar}
\end{document}